**Electric Vehicle Public Charging Equity Considerations: A Systematic Review**

**Boyou Chen**
ORCID: 0009-0007-6845-2467
Industrial and Manufacturing Systems Engineering Department
University of Michigan-Dearborn
4901 Evergreen Rd, Dearborn, MI, 48128, USA
Email: boyou@umich.edu

**Kaihan Zhang**
ORCID:  0009-0007-1502-571X
Cho Chun Shik Graduate School of Mobility
Korea Advanced Institute of Science and Technology
193 Munji–ro, Yuseong–gu, Daejeon, 34051, South Korea
Email: kaihn@kaist.ac.kr

**Austin Moore**
ORCID:  0009-0001-5205-046X
Industrial and Manufacturing Systems Engineering Department
University of Michigan-Dearborn
4901 Evergreen Rd, Dearborn, MI, 48128, USA
Email: austinmo@umich.edu

**Bochen Jia (Corresponding Author)**
ORCID:  0000-0002-0320-2789
Industrial and Manufacturing Systems Engineering Department
University of Michigan-Dearborn
4901 Evergreen Rd, Dearborn, MI, 48128, USA
Email: bochenj@umich.edu

**Mengqiu Cao**
ORCID:  0000-0001-8670-4735
Bartlett School of Environment, Energy and Resources
University College London
Gower Street, London, WC1E 6BT, United Kingdom
Email: mengqiu.cao@ucl.ac.uk

**Abstract**

Public electric vehicle (EV) charging infrastructure is crucial for accelerating EV adoption and reducing transportation emissions; however, disparities in infrastructure access have raised significant equity concerns. This systematic review synthesizes existing knowledge and identifies gaps regarding equity in EV public charging research. Following structured review protocols, 91 peer-reviewed studies from Scopus and Google Scholar were analyzed, focusing explicitly on equity considerations. The findings indicate that current research on EV public charging equity mainly adopted geographic information systems (GIS), network optimization, behavioral modeling, and hybrid analytical frameworks, yet lacks consistent normative frameworks for assessing equity outcomes. Equity assessments highlight four key dimensions: spatial accessibility, cost burdens, reliability and usability, and user awareness and trust. Socio-economic disparities, particularly income, housing tenure, and ethnicity, frequently exacerbate inequitable access, disproportionately disadvantaging low-income, renter, and minority populations. Additionally, infrastructure-specific choices, including charger reliability, strategic location, and pricing strategies, significantly influence adoption patterns and equity outcomes. However, existing literature primarily reflects North American, European, and Chinese contexts, revealing substantial geographical and methodological limitations. This review suggests the need for more robust normative evaluations of equity, comprehensive demographic data integration, and advanced methodological frameworks, thereby guiding targeted, inclusive, and context-sensitive infrastructure planning and policy interventions.

# 1. Introduction

Electric mobility has rapidly emerged as a cornerstone in the global endeavor to reduce carbon emissions and achieve more sustainable transportation networks. The International Energy Agency (IEA) highlights substantial global disparities in access to EV public charging, emphasizing the importance of equitable infrastructure to enable a widespread shift to electric mobility (IEA, 2024). Various national governments have intensified efforts to deploy robust EV public charging networks, allocating significant budgetary outlays and incentives for both charging infrastructure and advanced battery technologies. Notably, several European nations and China continue to pilot large-scale charging infrastructure initiatives, while in the United States, recent legislative instruments, alongside directives from agencies such as the U.S. Department of Transportation, aim to catalyze EV adoption through the expansion of public charging access.

Despite the accelerating policy momentum and investments, critics caution that uneven charging infrastructure rollout may reinforce or even exacerbate socio-economic and spatial inequities. Several studies indicate that the widespread deployment of strategically placed charging stations can mitigate range anxiety and encourage broader EV adoption (Niederle et al., 2021; Wolbertus et al., 2021). Early efforts in infrastructure expansion often focused on ensuring reliable coverage, with governments and private stakeholders installing chargers along major highways and in city centers to increase consumer confidence in EV usability on inter-city and intra-urban trips (Adenaw and Lienkamp, 2021; Seikh and Mandal, 2022). However, as EVs expand beyond affluent or early adopters into mainstream markets, researchers and policymakers increasingly recognize that simply achieving coverage is insufficient. Attention must also be paid to whether the infrastructure is evenly distributed, how it caters to different user groups, and how it can be scaled sustainably while meeting evolving mobility and energy demands (Namdeo et al., 2014; Pan et al., 2024). Research also shows that the lack of equitable and user-centric development could leave traditionally underserved communities, including rural and low-income populations, behind in the shift to electric mobility (Foley et al., 2020; Stajić et al., 2023). For instance, wide-ranging accessibility studies in dense urban areas have demonstrated that installing chargers near lucrative commercial centers or affluent residential districts is often more profitable for private operators, however, it frequently results in the unintentional underserving of neighborhoods where residents lack private home charging options, often lower-income communities or areas dominated by multi-unit dwellings. This disparity constitutes a critical equity issue, as those most reliant on public infrastructure face the greatest access barriers. Uneven charger distribution risks limiting EV adoption in these communities, potentially excluding them from the economic and environmental benefits of electric mobility and deepening existing socioeconomic divides (Falchetta & Noussan, 2021; Bhat et al., 2024b; Ding & Wu, 2025; Guo & Wang, 2023; Jha et al., 2025). Consequently, resolving issues of inclusivity and equitable distribution of charging stations has emerged as a focal challenge not only for municipalities, utilities, and regulators but also for EV manufacturers and prospective investors in charging services.

Beyond equity concerns, a second dimension involves the interplay of user behavior with charging infrastructure design. EV user behaviors can be broadly classified into five categories: routine-driven (regular commuting patterns) (Almaghrebi et al. 2019), convenience-oriented (minimal route deviation) (Jonas et al. 2023; Wang et al. 2019), economic (price sensitivity) (Kajanova et al. 2022; Visaria et al. 2022; Liu et al. 2022), risk-management (range anxiety mitigation) (Guo et al. 2018; Zatsarnaja et al. 2025), and time-sensitive (scheduling constraints) (Dominguez-Jimenez et al. 2020; Lei et al. 2022). These patterns directly influence charging decisions through distinct mechanisms. Routine-driven users favor charging locations integrated into daily routines, while convenience-oriented users prioritize minimal detours. Economically motivated users respond to pricing incentives and time-of-use rates, whereas risk-averse users charge preemptively even when unnecessary.

EV public charging decisions consequently hinge on a variety of factors, ranging from cost, time-of-day rates, charging speeds, and trip patterns, to psychological aspects such as range anxiety and perceived station reliability (Jonas et al., 2023; Shen et al., 2021). Evidence points to frequent mismatches between the available infrastructure and actual driver needs. For example, while many current deployments favor fast-charging corridors along highways, a substantial portion of EV owners may value slower but more conveniently located chargers, particularly near work sites or shopping malls. In this context, ignoring the heterogeneous behavior of EV users can diminish returns on large-scale infrastructure investments.

The impetus for this systematic review lies at the intersection of two pressing concerns: (1) ensuring equitable deployment of EV public charging networks, and (2) accounting for user behaviors and charging preferences in planning models. Although policy discussions around EV public charging infrastructure continue to expand, particularly in light of the United Nations' Sustainable Development Goals, extant research has only begun to address how equity, accessibility, and technological imperatives converge. Often, location-allocation models and operational frameworks for EV public charging largely assume homogeneous or idealized EV user charging behaviors, while socio-demographic insights remain under-integrated.

The review aims to answer the following research questions (RQs) to understand how literature measures, what factors matter, where, and with what effects for these factors in the context of EV public charging equity:

- RQ1: What are the main methodological frameworks and analytical approaches used to assess accessibility and equity of public charging infrastructure?

- RQ2: What are the key factors influencing accessibility and equity of EV public charging at the micro, socio-economic, infrastructure, and system levels?

- RQ3: How do the identified influencing factors affect the accessibility and equity of EV public charging across the geographic, socio-economic, and climate contexts?

Following this introduction, Section 2 elaborates on the systematic methodology employed to gather and analyze the relevant literature on EV public charging. Section 3 distills the findings into thematic clusters, summarizing existing methodologies adopted into public charging equity assessment, listing identified factors categorized into four groups that influence public charging equity, and discussing the implications for those factors' impacts. Section 4 synthesizes key insights, maps out policy-oriented strategies for equitable public charging expansion, and identifies important directions for future research, including the opportunities presented by advanced battery technologies, real-time pricing models, and cross-sector collaborations in mitigating disparities.

# 2. Methodology

## 2.1. Material Search and Inclusion Strategy

### *Resources Search and Identification*

We used a systematic review approach to identify and synthesize scholarly works related to EV public charging infrastructure. The database search was conducted in **Scopus** and **Google Scholar**, using sets of keywords that combined terms *"electric vehicle" OR "EV"* with *"public charging" OR "publicly accessible charging" OR "non-residential charging"* and relevant modifiers (*"charging infrastructure" OR "charging station"*). Additional keywords captured dimensions of equity and accessibility (*"disparity," "inclusi\*,"* "social justice"), user behavior (*"travel behavior," "charging behavior," "preference," "adoption"*), and socio-economic/demographic factors (workflow showing in **Figure 1**).

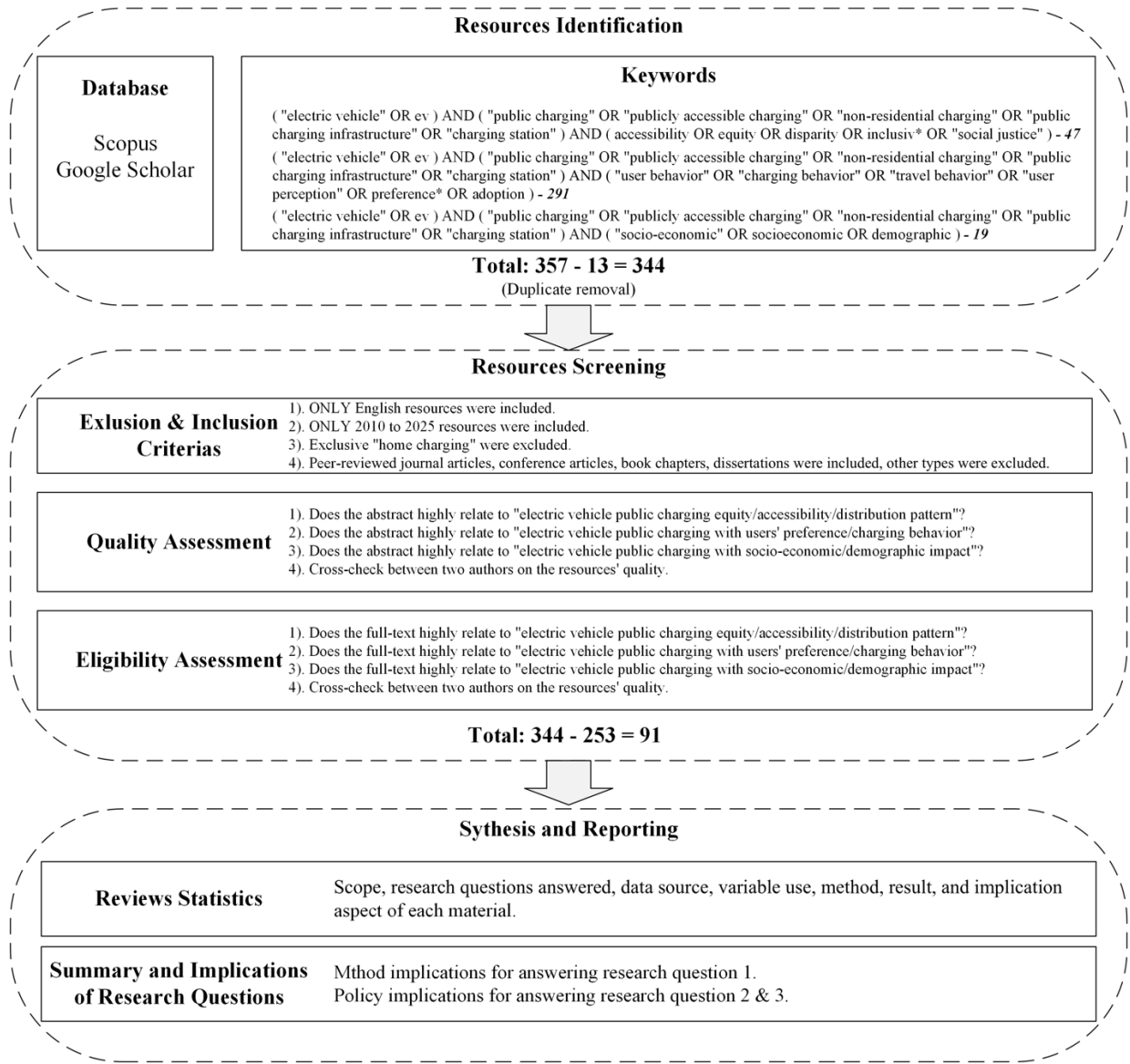

**Figure 1. PRISMA Search and Selection Workflow**

*Resources Screening, Quality, and Eligibility Assessment*

The search yielded a combined total of **357** references from both databases by using pre-defined keyword combinations. Then duplicate records (13 in total) were removed, resulting in **344** unique items. A set of exclusion and inclusion criteria was applied to screen those items, including: 1). Only English-language publications were retained. 2). Only studies published between 2010 and 2025 were considered. Large-scale public-charging roll-outs and the first peer-reviewed equity studies did not appear until the early 2010s, so earlier literature provides little relevant evidence. 3). Studies focusing exclusively on "home charging" were excluded. 4). Peer-reviewed journal articles, conference papers, book chapters, and dissertations were included, while excluding non-peer-reviewed sources such as reports, news articles, or editorials.

We screened titles and abstracts to confirm relevance to at least one of three key review focuses: 1) **Equity, accessibility, or distribution** aspects of public charging. 2) **User preference or charging behavior** associated with public charging. 3) **Socio-economic or demographic** impacts of public charging. Two authors independently conducted this screening to ensure quality and relevance, with disagreements resolved through discussion. Papers deemed relevant at the abstract level were subject to **full-text review** to verify their detailed treatment of the same three focus areas. Finally, each full-text was examined by two authors for final eligibility, again resolving any discrepancies via consensus. Through this iterative screening and assessment process (**see Figure 1**), we compiled the final set of studies for in-depth synthesis. The subsequent sections elaborate on the synthesis, interpretation, and policy implications derived from these selected materials. One thing worth noting is that to ensure no relevant materials were missed, backward tracking and forward tracking were further conducted for the selected 91 studies by reviewing their reference lists (backward tracking) and inspecting all articles that cite them in Web of Science (forward tracking) following defined inclusion and exclusion criteria; the results remained unchanged.

## 2.2. Review Statistics

A total of **91** relevant studies were included in this systematic review on equity and charging behaviors in EV public charging infrastructure. **Figure 2** provides summaries of the main statistical results, which are also discussed below. First, scholarly output has accelerated sharply since 2021: publications rose from 4 papers in 2020 to 13 in 2021, 16 in 2022, 15 in 2023, and 20 in 2024. The first quarter of 2025 has already published 13 studies, signaling sustained momentum and continued policy relevance. By counting the co-occurrence of keywords that appeared in 91 studies and mapping publication years, **Figure 2** illustrates an interesting trend that reflects the growing academic and policy interest in sustainable and equitable transportation solutions, particularly as EV equity becomes more mainstream. Because VOSviewer treats singular and plural tokens separately, both 'electric vehicle' and 'electric vehicles' appear as distinct nodes, although they designate the same research theme. Throughout the discussion, we treat them as one concept.

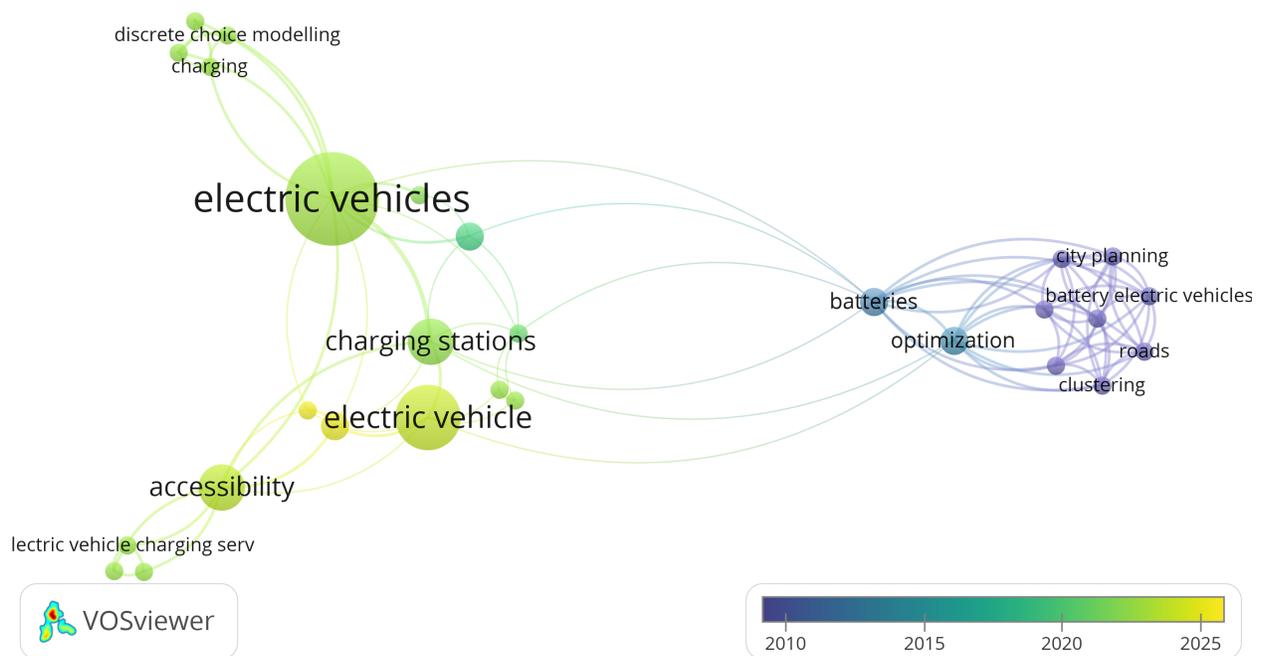

**Figure 2. Keywords' Co-occurrence Networks Among Observed Studies (Produced by VOSviewer)**

Regarding the study type, the majority of the reviewed papers (76 studies; 83.5%) were published as journal articles, while 15 studies (16.5%) appeared in conference proceedings. With regard to the type of data utilized, the analysis reveals that 52.7% of the studies relied on secondary datasets, whereas 38.5% employed primary empirical data collected through surveys, experiments, or direct measurements. Geographically, the studies covered a wide range of regions. China (23 studies, 25.27%), the United States (23 studies, 25.27%) were the most-analyzed countries, together accounting for almost half of all studies. India, Germany and the Netherlands followed (6 studies, 6.59%; 6 studies, 6.59%; 4 studies, 4.40%, respectively). A long tail of single- or double-study appearances includes Norway, New Zealand, Kuwait and several sub-national regions (for instance, Ontario, King County). 7 (7.69%) studies adopted a global theoretical scope untethered to real-world locations, signaling an emerging, but still modest, interest in transferable modelling frameworks. This pattern indicates that while the research interest is global, empirical field-based studies remain largely concentrated in high-income countries. The data sources used were highly diverse, encompassing survey-based data, public datasets, corporate data from electric utilities, social media-derived information, and synthesized data generated through simulation models. Survey data and public databases were among the most common sources, although the significant use of synthetic datasets in some studies points to ongoing challenges in accessing comprehensive real-world data.

The distribution of research questions discussed by the reviewed studies provides further insights. 58 studies were observed to primarily focus on behavioral, operational, usage aspects of EV public charging, making it the most prominent research focus. 49 studies were found to discuss public charging equity assessments and social dimensions. In comparison, 26 studies appeared to relate to public charging spatial distribution, while only 16 studies dealing with systemic barriers and policy implications, was the least addressed. Furthermore, an analysis of research focus co-occurrence revealed that operational aspects and equity dimensions are frequently studied together, suggesting an increasing integration of usage behavior analysis with social equity concerns. In contrast, infrastructure planning and systemic barriers are rarely discussed in combination, indicating a potential fragmentation in how infrastructure and policy challenges are conceptualized. Future research would benefit from more integrated frameworks that simultaneously address spatial planning, policy barriers, and equity outcomes.

# 3. Findings and Discussions

Based on 91 studies, this section first surveys the analytical lenses applied to charging equity (Section 3.1 answers RQ1), then synthesizes multi-level factors (Section 3.2 answers RQ2), and finally discusses how those determinants manifest across geographic, socio-economic, and climate contexts while highlighting evidence gaps (Section 3.3 answers RQ3).

## 3.1. Measurement of EV Public Charging Equity

### 3.1.1. Existing methodological frameworks

Within 91 selected studies, 52 were reviewed to have adopted a wide range of methodological frameworks to assess public charging infrastructure accessibility, each with distinct strengths and limitations. Based on (1) how the physical charging network, user demand, and policy levers are abstracted; (2) the mathematical or computational paradigm used to link inputs to outputs; and (3) the equity-related indicators it produces. **Table 1** provides a classification summary of existing frameworks proposed in previous literature. Individual studies may implement a framework with different datasets or solution techniques, but they remain comparable because they operate under the same overarching analytical architecture.

Specifically, Network-equilibrium and flow-capturing frameworks treat chargers as nodes on coupled road–power networks and solve mathematical-programming or optimization problems to reach system-wide equilibrium, thereby reproducing routing constraints and queuing dynamics (Wang et al., 2019; Xie & Lin, 2021; Qiao et al., 2023). Spatial-accessibility frameworks adopt GIS coverage indices, floating-catchment areas and kernel-density estimates to juxtapose charger supply with population demand, producing intuitive hotspot maps for planners even though their outputs can be distorted in low-density regions and are sensitive to radius or travel-time thresholds (Bansal et al., 2021; Niederle et al., 2021; Leng et al., 2023; Oluwajana et al., 2023; Zhang & Fan, 2025). From user perspective, behavioral-decision frameworks incorporate heterogeneous user preferences, including value of time, range anxiety, and socio demographics, through discrete choice, Bayesian, or AI estimators, yielding equity-relevant insights at the cost of large, representative survey requirements and potential stated-preference bias (Li et al., 2018; Ahmed et al., 2021; Bhat et al., 2024; Huang et al., 2024; Jiang et al., 2024; Li et al., 2025).

Finally, integrated or hybrid frameworks purposely fuse two or more of the foregoing approaches, for example, Analytic Hierarchy Process (AHP) – entropy – Technique for Order Preference by Similarity to Ideal Solution (TOPSIS) multi-criteria ranking embedded in an equilibrium model or GIS hotspot detection feeding a system-dynamics loop to reconcile engineering precision with social realism, though such syntheses raise data demands and computational complexity (Shen et al., 2021; Patil et al., 2023; Guo et al., 2025).

**Table 1. Summary of Existing Adopted Frameworks to Assess Equity in Public Charging Infrastructures.**

| Framework | Main idea | Outputs | Strengths | Limitations | Relevant Study Count (%) |
|---|---|---|---|---|---|
| **Network-equilibrium & flow-capturing** | Treat charging stations as nodes on a multimodal network; jointly optimize travel cost and charging demand. | Optimal station siting, equilibrium flows. | Captures route substitution and queuing effects. | High data and parameterization burden; equilibrium assumptions may break down under stochastic demand. | 8 (15.38%) |
| **Spatial-accessibility** | Measure geographic reach (for instance, coverage radius, 2SFCA, kernel density) relative to population or vehicle stock. | Accessibility scores, hotspot maps. | Intuitive, GIS friendly; works with sparse data. | May over- or under-estimate access in rural/low-pop density areas; ignores temporal variation. | 16 (30.77%) |
| **Behavioral-decision** | Embed user heterogeneity (value of time, range anxiety, socio-demographics) in choice or game-theoretic models. | Choice probabilities, elasticities, equity impacts. | Captures distributional effects; aligns with survey evidence. | Sensitive to survey bias & stated-preference artefacts; heavy parameterization. | 20 (38.46%) |
| **Hybrid frameworks** | Combine two or more of the above (for instance, equilibrium + behavioral logit, or GIS kernel density feeding an system dynamic adoption loop). | Multi-scale KPIs (coverage, queuing delay, equity index). | Bridges technical & socio-behavioral lenses; better policy realism. | Integration increases data demands and computational complexity. | 8 (15.38%) |

Overall, these frameworks have broadened the empirical and conceptual base for evaluating EV public charging accessibility, yet network-equilibrium remains numerically smaller, largely due to higher demand of rich traffic, power-flow, and socio-economic data, which is similar to the adoption of hybrid frameworks. The choice of these frameworks therefore depends on the planning objective and data environment: long-range corridor design favors equilibrium models; cities lacking detailed trip data may start with GIS coverage indices; equity audits or tariff design call for behavioral models; and large public-private programs increasingly adopt hybrids to capture grid constraints and demographic heterogeneity. Framing the approaches along this strategic–tactical–operational spectrum clarifies that they are orthogonal tools in a common toolbox rather than rungs on a maturity ladder. To further investigate what are the current potential methodology gap in measuring EV public charging equity, detailed analytical approaches are discussed for both engagement in 52 studies related to frameworks and adoption frequency among 91 studies.

### 3.1.2. How do analytical approaches engage in existing frameworks

Analytical approaches to assess public charging accessibility through network-optimization logic deploy mathematical programming or flow-capturing models to decide where and how large stations should be. Corridor-scale studies use mixed-integer and flow-capturing formulations to minimize system-wide travel cost or traveler "inconvenience" (Xie and Lin, 2021; Qiao et al., 2023). More behaviorally rich variants add Bayesian or car-following sub-modules (Li et al., 2018; Guo et al., 2025) and even incorporate equity or dual objectives of efficiency and fairness (Liu et al., 2025). Their prescriptive power is high, yet all remain sensitive to congestion-queuing parameters and become computationally prohibitive once national networks or stochastic demand scenarios are introduced (Ip et al., 2010; Wang et al., 2019; Wu and Ribberink, 2020). Meta-heuristic hybrids (genetic algorithms with mathematical programs, for instance) alleviate but do not eliminate scale and calibration issues (Ouyang and Xu, 2022).

A second stream relies on spatial-statistical and GIS techniques to portray how chargers are distributed relative to people, trips, or deprivation scores. Kernel-density estimation, floating-catchment areas, and G2SFCA measures dominate this family, producing intuitive equity maps but often overstating access in low-density tracts (Niederle et al., 2021; Li et al., 2022). Cluster-detection algorithms such as DBSCAN, hierarchical, and K-Prototype approaches sharpen hotspot detection and reveal socio-spatial disparities (Richard et al., 2022; Leng et al., 2023; Ding and Wu, 2025). Multi-criteria decision-making (AHP, entropy weighting, improved TOPSIS) is regularly coupled with GIS layers to rank candidate sites (Shen et al., 2021; Seikh and Mandal, 2022) or to blend efficiency and equity targets (Esmaili et al., 2024; Nazari-Heris et al., 2022). Still, these approaches hinge on user-defined buffer radii and survey-derived demand weights, making cross-city transferability weak (Sheng et al., 2025; Namdeo et al., 2014).

Machine learning (ML) pipelines wrapped in micro-services forecast temporal demand and feed siting heuristics (Thorve et al., 2023; Ahmed et al., 2021), whereas spatial–temporal load-forecasting models link trip chains, Dijkstra routing and coordinated slow/fast-charge scheduling to anticipate sub-hourly load bursts (Tang et al., 2021). Stock-and-flow system dynamics or multi-stakeholder negotiations embed policy feedbacks and scheduling constraints (Karmaker et al., 2023; Patil et al., 2023), and several studies now propose analytical templates that explicitly fold equity metrics into broader accessibility goals (Cai et al., 2024; Li et al., 2018). These integrated approaches promise greater realism but demand high-resolution mobility traces, charger-load data and interdisciplinary calibration, placing large data and computational burdens on researchers.

**Table 2** further summarizes six detailed approaches that were frequently used in observed studies. Optimization and mathematical programming models lead the pack with 10 papers (11 %), underscoring the engineering focus on siting and capacity sizing. Regression and other econometric tools and spatial-statistical / GIS techniques follow closely, each with nine studies (9.9 %), and together anchor most empirical work on charger use and socio-spatial patterns. Clustering and other unsupervised methods (eight studies; 8.8 %) help uncover latent demand hotspots, while simulation approaches, for instance, trip-chain, agent-based, or power-flow, appear with the same frequency and capture time-varying vehicle-grid interaction. Finally, structured surveys and stated-preference experiments (five studies; 5.5 %) and a small but growing group of machine-learning pipelines (four explicit ML-first articles; 4.4 %) add behavioral depth and predictive power.

Each approach brings clear strengths—precise optimization, transparent regression, intuitive GIS mapping, data-driven clustering, dynamic simulation, or direct behavioral insight—but they also share two concrete gaps. First, only a handful of studies compare their model outputs with observed charger utilization or track equity outcomes over time, so external validity remains limited. Second, methods are still used in silos: optimization models rarely incorporate the rich user heterogeneity revealed by surveys or ML forecasts, while spatial and behavioral studies seldom respect the physical constraints captured in network optimization. Moving forward, researchers will need publicly available, high-resolution travel and load datasets and modular workflows that let optimization engines, GIS layers, and behavioral models exchange data efficiently and transparently. Such integration would allow planners to test whether a site that is "optimal" on paper also performs under real travel patterns and delivers fair access across communities.

**Table 2. Frequently Used Approaches in Assessing Public Charging Infrastructure Accessibility**

| Approach class (keywords) | Studies (N = 91) | % of corpus | Cum. (%) | Strengths | Assumptions/Cautions |
|---|---|---|---|---|---|
| **Optimization** | 10 | 11 % | 11.00% | Precise siting & sizing decisions; handles constraints explicitly. | Parameter-sensitive; solution time grows quickly with network size. |
| **Regression** | 9 | 9.9 % | 20.90% | Quantifies utilization drivers; easy statistical inference. | Causal interpretation tenuous if confounders omitted. |
| **Spatial statistics** | 9 | 9.9 % | 30.80% | Hotspot identification, clustering, Moran's I for equity patterns. | Results hinge on distance thresholds and population weighting. |
| **Clustering & unsupervised learning** | 8 | 8.8 % | 39.60% | Reveals latent usage archetypes without pre-defined classes. | Sensitive to feature scaling; cluster meaning must be interpreted post-hoc. |
| **Simulation (trip-chain, agent-based, power-flow)** | 8 | 8.8 % | 48.40% | Captures temporal dynamics and vehicle–grid interactions. | Requires granular Origin-Destination (OD), charger and |

| Approach class (keywords) | Studies (N = 91) | % of corpus | Cum. (%) | Strengths | Assumptions/Cautions |
|---|---|---|---|---|---|
| | | | | | battery data that are rarely public. |
| Structured surveys | 5 | 5.5 % | 53.90% | Direct insight into user willingness-to-pay, equity perceptions. | Sampling bias; stated vs. revealed behavior gap. |
| Advanced ML | 4* | 4.4 % | 58.30% | High predictive accuracy for demand and dwell time. | Data-hungry; black-box models hinder policy transparency. |

*\* ML count reflects only explicit ML-first studies; several optimization papers embed ML modules and are counted above.*
*Note: "% of corpus" refers to (number of studies in that row ÷ 91 total studies) × 100, "Cum. (%)" refers to the cumulative percentage.*

## 3.2. Multi-level Influencing Factors of EV Public Charging Equity

Existing studies have adopted various analytical approaches that involve considering factors into five types of frameworks, and identified multiple factors that can significantly impact public charging infrastructure accessibility and equity across regions and demographic groups.

### 3.2.1. Micro-Level Factors: User Charging Behavior, Preference, EV Performance

At the micro scale, an EV user's observable public charging behavior is the outcome of two interacting forces: (1) the technical constraints of the vehicle (battery size, charging speed) and (2) the EV user's preferences and psychological costs (for instance, range anxiety, value of time).

User preferences reflect the relative importance individuals place on different charging attributes, with cost sensitivity and convenience being central to infrastructure utilization. Discrete choice models consistently show that users value proximity, speed, and reliability in public charging options (Bansal et al., 2021; Huang et al., 2024; Nazari-Heris et al., 2022; Srivastava et al., 2022). Embedding psychological cost terms changes optimal locations and lowers detour kilometers in Sioux Falls (Guo et al., 2018). EV performance characteristics refer to the technical specifications of the EV itself, such as battery range and charging speed, which directly influence the frequency and nature of charging needs. Owners of smaller-battery EVs often require more frequent public charging access (Ding and Wu, 2025; Loh and Noland, 2024). Integrated models using real-time pricing and mobility patterns illustrate how dynamic pricing structures can interact with user behavior and infrastructure demand (Kajanova et al., 2022; Lei et al., 2022; S. Zhou et al., 2021).

Charging behaviors then encompass the patterns and decisions EV users make regarding when, where, and how often they charge their EVs. These behaviors are shaped by variables such as the state of charge (SOC), proximity to charging stations, and trip characteristics. Long-distance travelers prioritize fast charging and reduced waiting times (Ge and MacKenzie, 2022; Leng et al., 2023; Solvi Hoen et al., 2023), whereas urban users exhibit more varied charging patterns based on location types, such as workplaces and shopping centers, as well as seasonal factors (Almaghrebi et al., 2019; Jonas et al., 2023; Khaleghikarahrodi and Macht, 2023; Dominguez-Jimenez et al., 2020; Dost et al., 2018). Data-mining work on 189 k Illinois ChargePoint sessions confirms that morning, multifamily or commercial sites and low start-SOC dominate public use, while workplace sessions are shorter on weekdays (Siddique et al., 2022). Ride-hailing traces from Nanjing show clear private vs. commercial patterns; a Cumulative-Prospect-Theory model reveals risk-seeking station choice that cuts waiting time by>10 % (Xing et al., 2021). Seasonality also matters: winter demand is lowest and autumn highest, with random-forest forecasts yielding MAPE 0.08 % (Dominguez-Jimenez et al., 2020). Early work in Amsterdam derived a user-type taxonomy (residents, commuters, taxis, car-sharing, etc.) with distinct temporal-load profiles, providing a foundation for utilization forecasting (Helmus and Van Den Hoed, 2015). Behavioral studies further reveal that range anxiety, trip purpose, and daily routines are key drivers of charging decisions, with users often preferring chargers conveniently located along daily travel routes (Jha et al., 2025; Li et al., 2025; Wang et al., 2019). While these micro-level forces shape who seeks public chargers and under what circumstances, the extent to which that demand is actually satisfied depends on socio-economic and demographic factors.

### 3.2.2. Socio-economic and Demographic Influences

Socioeconomic and demographic factors, including income, education, and housing status, have a profound influence on EV adoption and infrastructure accessibility. Higher-income and better-educated populations typically enjoy greater access to public charging facilities, leading to pronounced equity gaps (Bansal et al., 2021; Bhat et al., 2024a; Falchetta and Noussan, 2021; Huang et al., 2025; Oluwajana et al., 2023). Recent spatial–statistical analyses across ten Chinese cities found significant intra-city inequity in public charging station access (Li et al., 2022). Home ownership and housing type also determine the ability to charge at home, with renters and individuals living in multi-unit dwellings more dependent on public charging networks (Bhat et al., 2024b; Xu et al., 2024; Zhang and Zhang, 2022). Furthermore, existing infrastructure often clusters in affluent or commercial areas, exacerbating accessibility barriers for marginalized groups (Gazmeh et al., 2024; Zatsarnaja et al., 2025; Zhang and Fan, 2025). Studies have observed that certain nationalities or ethnic groups dominate EV usage in some contexts, indicating uneven participation in the EV transition (Banna et al., 2023; Li et al., 2017; Pan et al., 2024; Stajić et al., 2023). These socio-demographic filters determine who must rely on public charging; the next section (3.2.3) examines whether the existing network's availability, siting and design translate that latent demand into real accessibility.

### 3.2.3. Infrastructure Supply & Operations: Availability, Siting, Design, and Pricing

Research consistently shows that the availability, strategic placement, and thoughtful design of EV public charging infrastructure play a crucial role in influencing EV adoption rates, user awareness, and utilization patterns across diverse regions and demographic groups. Strategic infrastructure deployment reduces range anxiety and enhances convenience, thereby increasing EV adoption. Spatial analyses in New Zealand (Selena Sheng et al., 2022) and optimization models in China (Ouyang and Xu, 2022) demonstrate the importance of neighborhood effects, while a bi-level siting model that embeds a range-anxiety cost term shifts the optimal fast-charger set and cuts detour distances (Guo et al., 2018). Large-scale trace mining in Nanjing further shows that risk-seeking drivers pick stations that minimize perceived wait time; a Cumulative-Prospect-Theory engine trims waiting costs by >10 % and illustrates how behavior-aware siting can boost perceived accessibility (Xing et al., 2021). Similarly, areas with denser and more reliable charging networks consistently show higher EV adoption, as accessible infrastructure improves user confidence and lowers barriers to entry (Bhat et al., 2024a; Oluwajela et al., 2023; Wang et al., 2019; Zhang and Zhang, 2022).

Placement strategies that prioritize high-traffic, visible, and convenient locations have been shown to boost user awareness and the perceived viability of EVs for everyday travel (Falchetta and Noussan, 2021; Li et al., 2025; Zhang and Fan, 2025). For instance, clustered stations in Amsterdam (Wolbertus et al., 2021) and fast-charging hubs along U.S. corridors (Xie and Lin, 2021) enhance accessibility and foster adoption through reciprocal effects, where increased infrastructure availability encourages more users. A coordinated load-forecasting framework that links trip-chain demand with home, destination, and en-route charging helps operators proactively avoid local overloads (Tang et al., 2021). Theoretical simulations suggest that access to both home and public charging amplifies market share through indirect network effects, highlighting the synergistic benefits of a combined infrastructure strategy (Kishi et al., 2023). However, current deployments often favor urban and affluent areas, leaving rural and underserved communities with limited access to charging, which restricts adoption potential and perpetuates regional disparities (Bhat et al., 2024b; Ding and Wu, 2025; Gazmeh et al., 2024).

Infrastructure design features, including charger type (Level 2 versus DC fast charging), charging speed, pricing models, and user interface design, have significant impacts on utilization and user satisfaction (Bansal et al., 2021; Huang et al., 2024; Srivastava et al., 2022; Zatsarnaja et al., 2025). Studies in Norway (Solvi Hoen et al., 2023) and Slovakia (Kajanova et al., 2022) reveal diverse user preferences for fast charging and smart charging incentives, which influence how infrastructure is perceived and used. Conversely, experiences of "Charge Point Trauma" in the United Kingdom (Chamberlain and Al Majeed, 2021) show that issues such as charger operability, reliability, and complex payment systems can hinder user satisfaction and discourage adoption. High-speed chargers placed along major travel corridors support long-distance EV mobility, especially for drivers lacking reliable home charging options (Li et al., 2023; Nazari-Heris et al., 2022). In contrast, poorly maintained or confusingly designed stations deter usage, particularly among first-time EV drivers or those less familiar with EV technologies (Chang and Oh, 2024; Huang et al., 2025; Xu et al., 2024). Beyond physical design, cooperative deep-reinforcement-learning schemes for multi-station networks raise total profit by $\approx$ 24 % and smooth loads compared with independent pricing, demonstrating how adaptive tariffs can both entice users and protect the grid (Liu et al., 2022). Similar coordinated-behavior load-forecasting models improve network-operation forecasts (Tang et al., 2021).

Regional variations also play a crucial role in shaping outcomes. In Canada, the "EV Duck Curve" illustrates challenges related to grid stability caused by the clustering of public fast-charger demand (Jonas et al., 2023), while infrastructure scarcity in Kuwait limits EV adoption to wealthier nationals (Banna et al., 2023). Cold climates further exacerbate charging needs by reducing battery efficiency, requiring more frequent charging and impacting infrastructure demand (Chang and Oh, 2024, 2024; Jha et al., 2025). AI-driven design improvements (Ahmed et al., 2021) have enhanced user confidence globally, suggesting the potential for technology-driven solutions to improve infrastructure utilization. Despite these advances, several limitations remain, including a reliance on hypothetical scenarios (Ge and MacKenzie, 2022), small sample sizes (Krause et al., 2018), and region-specific data (Wei et al., 2022), which highlight the need for broader empirical validation across diverse socioeconomic and geographic contexts.

Collectively, these supply-side choices ripple outward, including driving EV adoption patterns, altering peak grid loads, influencing local economic activity, and reshaping emissions footprints. The system-level outcomes are examined next in section 3.2.4.

### 3.2.4. System-Level Impacts: Adoption, Grid Loads, Economic & Environmental Outcomes

The socio-economic and environmental impacts of EV public charging infrastructure have been explored through various frameworks and across different geographical contexts. Public charging infrastructure can stimulate local economic growth by attracting businesses, creating new job opportunities, and increasing property values near station locations (Bhat et al., 2024b; Wang et al., 2019; Xu et al., 2024; Y. Zhou et al., 2021). However, these benefits are often concentrated in wealthier urban areas, thereby exacerbating existing socio-economic disparities (Ding and Wu, 2025; Gazmeh et al., 2024; Huang et al., 2025). Regions with limited infrastructure investment face reduced access to economic opportunities, emphasizing the need for equity-focused planning and deployment strategies (Silva et al., 2024; Zhang and Fan, 2025).

On the environmental side, expanding public charging availability encourages EV adoption, leading to reductions in greenhouse gas emissions and urban air pollution (Bian et al., 2025; Oluwajana et al., 2023; Zhang and Zhang, 2022). However, the net environmental benefit of public EV infrastructure largely depends on the energy generation mix. Regions powered by renewable energy sources realize greater emission reductions, whereas areas still reliant on fossil fuels may see more limited environmental gains (Jha et al., 2025, 2025; Korkas et al., 2024). Climatic conditions further influence infrastructure performance and outcomes; colder climates, for example, reduce battery efficiency, necessitating more frequent and prolonged charging, which can stress local electricity grids and potentially increase emissions if clean energy is not available (Souto et al., 2024; Visaria et al., 2022; Zhou et al., 2025).

Several studies highlight emerging concerns about the resilience of EV public charging networks in the face of extreme weather events associated with climate change, which can disrupt charging availability and reliability (Huang et al., 2024; Li et al., 2023). Broader social impacts include shifts in transportation habits, land use changes, and urban dynamics, which require integrated and flexible planning approaches (Li et al., 2024; Liu et al., 2025, 2024; Roy and Law, 2022). Tailored regional strategies, such as incentivizing rural charging deployments or integrating charging stations with renewable microgrids, have been shown to promote more equitable and sustainable outcomes (Amezquita et al., 2024; Jonas et al., 2025; Sheng et al., 2025; Zatsarnaja et al., 2025).

Specific modeling efforts, such as a study using machine learning and microservices-oriented architecture, have demonstrated the potential of smart grid-integrated EV adoption and charging station planning to enhance energy sustainability, although these models often rely on synthetic data, limiting empirical applicability (Thorve et al., 2023). Similarly, Canadian studies reveal that EV public charging can contribute to new peak electricity loads, posing challenges to grid stability under seasonal variations, though socio-demographic and spatial data gaps remain a limitation (Jonas et al., 2023). A global panel data study covering 14 countries found that renewable energy availability, education, and population density, rather than GDP alone, influence EV demand (Li et al., 2017), though the study did not account for micro-level socio-economic factors.

Without deliberate and inclusive planning, public EV infrastructure risks reinforcing environmental injustices by disproportionately benefiting already advantaged populations (Falchetta and Noussan, 2021; Guo and Wang, 2023; Neumeier and Osigus, 2024). Innovative policy measures, such as dynamic pricing, public-private partnerships, and community-led planning, are proposed to maximize both the socio-economic and environmental benefits of EV public charging networks (Calvo-Jurado et al., 2024; Deb et al., 2020; Karmaker et al., 2023; Li et al., 2022; Wang et al., 2023). Overall, while EV public charging infrastructure offers substantial societal and environmental benefits, realizing these benefits equitably requires context-sensitive, inclusive, and forward-looking strategies that address regional and climatic differences.

Despite these insights, the effectiveness of the current charging infrastructure is often compromised by overly simplified planning approaches that fail to adequately account for dynamic user behaviors and regional variations (Namdeo et al., 2014; Patil et al., 2023). Socioeconomic conditions, particularly income, education, and race, interact with geographic factors, resulting in rural, minority, and low-income urban areas facing notably lower charger density (Esmaili et al., 2024; Li et al., 2023; Oluwajana et al., 2023; Zhou et al., 2025). Surveys suggest that minority and lower-income users have lower awareness and trust toward public charging reliability, often feeling marginalized in the shift to electric mobility (Deb et al., 2020; Du et al., 2024; Jha et al., 2025; Peng et al., 2025). Finally, policy and planning practices further complicate equity outcomes. Site selection processes frequently prioritize areas with higher expected utilization, inadvertently reinforcing accessibility advantages for affluent communities (Calvo-Jurado et al., 2024; Guo et al., 2025; Roy and Law, 2022). However, several studies highlight that targeted awareness campaigns and inclusive planning strategies can improve equity outcomes by addressing the needs and barriers faced by diverse user groups (Amezquita et al., 2024; Chang and Oh, 2024, 2024; Guo and Wang, 2023; Srivastava et al., 2022).

## 3.3 Effects of Identified Factors Across Diverse Contexts

Recent equity in transport and energy stresses that public charging justice is multi-dimensional; a single metric such as distance to the nearest charger cannot capture the full distribution of benefits and burdens. Building on factors identified from existing studies (recall section 3.2), we evaluate each factor against four dimensions that appear in the literature:

1. **Spatial access (SA).** Geographic equity asks whether chargers are located close enough for routine use by all groups. GIS studies link higher station density to shorter detours and faster adoption, while highlighting rural and minority "charging deserts" (Li et al., 2022; Falchetta and Noussan, 2021; Jiang et al., 2024).
2. **Cost burden (CB).** Even where chargers exist, users face heterogeneous out-of-pocket costs, including session fees, detour fuel, or peak-period tariffs. Income-stratified analyses show that low-income and renter households pay a higher share of disposable income for public charging (Bansal et al., 2021; Bhat et al., 2024b).
3. **Reliability & usability (RU).** Equity also hinges on whether chargers are operational, fast, and easy to use. Field audits reveal that a quarter of public DC fast chargers may be non-functional at a given time, disproportionately discouraging first-time and price-sensitive users (Chamberlain and Al Majeed, 2021).
4. **Awareness & trust (AT).** Social-psychological access includes knowledge of charger locations and confidence in their reliability. Surveys find lower familiarity and trust among minority and lower-income drivers, compounding physical and financial barriers (Deb et al., 2020; Du et al., 2024).

By considering the effects of factors on these four dimensions, EV public charging equity could be comprehensively evaluated from physical reachability (SA), direct or indirect user expenditures (CB), functional availability of charging infrastructures (RU), and knowledge of charger locations as well as confidence in their operability (AT), and then further identify potential evidence gaps in geographic, socio-economic, and climate contexts.

**Table 3** links each single factor identified in Section 3.2 to these four dimensions, indicating the direction of influence (↑ improves, ↓ worsens) and illustrating how the same variable can widen or narrow inequities depending on geographic or socio-economic context.

**Table 3  Factor Evidence on How Each Impact Four Equity Dimensions of EV Public Charging**

*SA = spatial access    CB = cost burden    RU = reliability & usability    AT = awareness / trust.*
*Arrows show direction: ↑ = increases / improves, ↓ = decreases / worsens, — = no observed impact.*

| Factor group | factor | SA | SA Impact | CB | CB Impact | RU | RU Impact | AT | AT Impact | Typical context | References |
|---|---|---|---|---|---|---|---|---|---|---|---|
| Micro-level | Range anxiety | ↓ | Users restrict search radius, choose nearest chargers | ↑ | Pay premium to avoid detours | — — | | ↓ | Lower trust where sites are few | Sioux Falls USA; national surveys | Guo et al., 2018; Jha et al., 2025; Li et al., 2025; Wang et al., 2019 |
| | Battery size | ↑ | Smaller packs require denser network | ↑ | More frequent paid top-ups per km | — — | | — — | | Entry-level EVs, CN & US | Ding & Wu, 2025; Loh & Noland, 2024 |
| | Trip purpose (work / retail / long-haul) | ↓ | Workplace sessions cluster at offices; retail at malls | — — | | — — | | ↑ | Regular commuters learn "home" sites | Amsterdam NL; Illinois US; Nanjing CN | Almaghrebi et al., 2019; Jonas et al., 2023; Siddique et al., 2022; Xing et al., 2021 |
| Socio-economic & demographic | Income level | ↑ | High-income areas host more chargers | ↓ | Higher incomes absorb fees; low incomes face ↑ burden | — — | | ↑ | Affluent users report more trust | 10 Chinese cities; US metros | Bansal et al., 2021; Bhat et al., 2024a; Falchetta & Noussan, 2021; Huang et al., 2025; Li et al., 2022; Oluwajana et al., 2023 |
| | Housing type (renters, MUDs) | ↓ | Renters & MUD residents rely on curbside public chargers | ↑ | Regular public use raises monthly spend | — — | | — — | | US multifamily; CN apartments | Bhat et al., 2024b; Xu et al., 2024; Zhang & Zhang, 2022 |
| | Ethnicity / nationality | | Minority districts have fewer stations | ↑ | Longer detours raise cost | — — | | ↓ | Lower awareness & trust | Kuwait; minority areas US | Banna et al., 2023; Li et al., 2017; Pan et al., 2024; Stajić et al., 2023; Deb et al., 2020; Du et al., 2024; Peng et al., 2025 |
| Infrastructure supply & | Network density (stations / km²) | ↑ | Higher density shortens average distance | — — | | — — | | — — | | NZ rollout; CN megacities | Selena Sheng et al., 2022; Ouyang & Xu, 2022; Bhat et al., 2024a; Oluwajana et al., 2023; Wang et al., 2019; Zhang & Zhang, 2022 |
| operations | Siting bias to affluent / high-traffic areas | ↓ | Rural & low-income zones left sparse | — — | | — — | | ↑ | Visibility boosts perceived access in wealthy zones | Amsterdam NL; US corridors | Falchetta & Noussan, 2021; Li et al., 2025; Zhang & Fan, 2025; Wolbertus et al., 2021; Xie & Lin, 2021 |
| | Charger reliability / operability | — — | | ↑ | Extra fuel / time when units fail | ↓ | Faults, slow payment apps | ↓ | Repeat users lose confidence | UK public audit | Chamberlain & Al Majeed, 2021 |
| | Dynamic pricing / smart scheduling | — — | | ↓ | Off-peak tariffs cut bills; peak rates ↑ burden | ↑ | Load balancing shortens queues | — — | | CN multi-station DRL studies | Liu et al., 2022; Tang et al., 2021; Kajanova et al., 2022; Lei et al., 2022; Zhou et al., 2021 |
| | Charger type (DCFC vs Level 2) | ↑ | DCFC extends viable trip range | ↑ | Higher per-kWh fees at DCFC | | Faster sessions; Level 2 is slower | — — | | Norway highways; Slovakia incentives; US | Bansal et al., 2021; Huang et al., 2024; Srivastava et al., 2022; Zatsarnaja et al., 2025; Solvi Hoen et al., 2023; Kajanova et al., 2022; Nazari-Heris et al., 2022 |
| System-level & climate | Cold-climate battery loss | ↓ | More stops needed in winter | ↑ | Extra energy & session fees | — — | | — — | | Nordic & cold US states | Chang & Oh, 2024; Jha et al., 2025; Souto et al., 2024; Visaria et al., 2022; Zhou et al., 2025 |
| | Peak-load / "EV duck curve" | — — | | ↑ | Peak tariffs or demand charges | ↓ | Voltage sag slows charging | — — | | Canada winter peaks | Jonas et al., 2023; Thorve et al., 2023 |

Table 3 demonstrates that public EV-charging equity is multidimensional, influenced by diverse factors across spatial access, cost burden, reliability and usability, and awareness and trust. Micro-level determinants, such as range anxiety and battery size, significantly affect users' reliance on public chargers, often driving increased financial burdens due to more frequent usage or longer detours. Additionally, daily trip purposes such as commuting or retail influence users' habitual charger selection and familiarity, potentially reinforcing existing spatial disparities.

Socio-economic and demographic factors further highlight stark equity issues. Income disparities frequently result in higher charger density in affluent areas, leaving lower-income communities underserved. Renters or individuals living in multi-unit dwellings disproportionately depend on public chargers, bearing higher cumulative costs and experiencing greater inconvenience. Ethnicity and nationality also emerge as critical factors, with marginalized groups often confronting compounded accessibility and trust barriers due to fewer chargers and lower perceived reliability.

Infrastructure supply and operational choices substantially shape user experiences. Higher charger density enhances spatial accessibility, reducing charging deserts, yet placement strategies favoring affluent or high-traffic zones frequently exacerbate inequities. Issues of charger reliability significantly impede usability, notably affecting first-time and financially constrained users, while dynamic pricing and smart scheduling can alleviate cost burdens but require careful implementation to avoid disadvantaging specific user groups.

Finally, broader system-level and climatic considerations have substantial impacts on equity outcomes. Cold climates and increased peak-load demands necessitate more frequent public charging, raising costs and exacerbating inequities in regions without robust infrastructure support. These findings underscore the necessity of comprehensive and integrated planning, emphasizing regional sensitivity, targeted infrastructure deployment, and proactive engagement with underserved communities to effectively address the multidimensional nature of EV public charging equity.

## 4. Conclusions

This systematic review synthesizes extensive findings from 91 peer-reviewed studies examining equity considerations in EV public charging infrastructure across North America, Europe, and Asia. The review points out the emerging trend in EV public charging equity research in recent years, critically compares major methodological frameworks and analytical methods used in prior research, and highlights essential factors influencing equitable access and adoption of EV public charging infrastructure.

This review offers several significant observations to guide future research. Firstly, this review classifies existing theoretical frameworks into four distinct groups demonstrated in **Figure 3**. A significant proportion of adopting spatial and user-related frameworks is observed (totally 36 studies, accounting for 69.23%), yet few studies have measured EV public charging equity from both spatial and behavioral perspectives by adopting hybrid framework due to data availability challenges. Findings from the analytical approaches support this distribution. Network-optimization work, which typically mixed-integer or bilevel formulations shows that adding simple equity constraints can raise system cost modestly yet cuts low-access-tract travel distance and wait time by one-third or more. Spatial-accessibility studies, grounded in kernel-density, 2SFCA or KDE-plus-Gini indices, repeatedly flag "charging deserts" in rural and low-income tracts even when regional charger counts appear adequate. Behavioral-decision most often involved multinomial logit or latent-class choice, confirm that income, dwelling type and range anxiety jointly outweigh distance in predicting public-charger uptake. Early hybrid experiments that link agent-based demand with power-flow simulation indicate that fast-charger clustering without demand response worsens both grid voltage sag and access inequity. Since each framework isolates only one side of the problem, including flows, geography, or user choice, comparative tests on a common dataset using standardized equity indicators are now essential. Such head-to-head evaluations will reveal where each framework excels, expose trade-offs, and guide the design of flexible, context-aware planning tools.

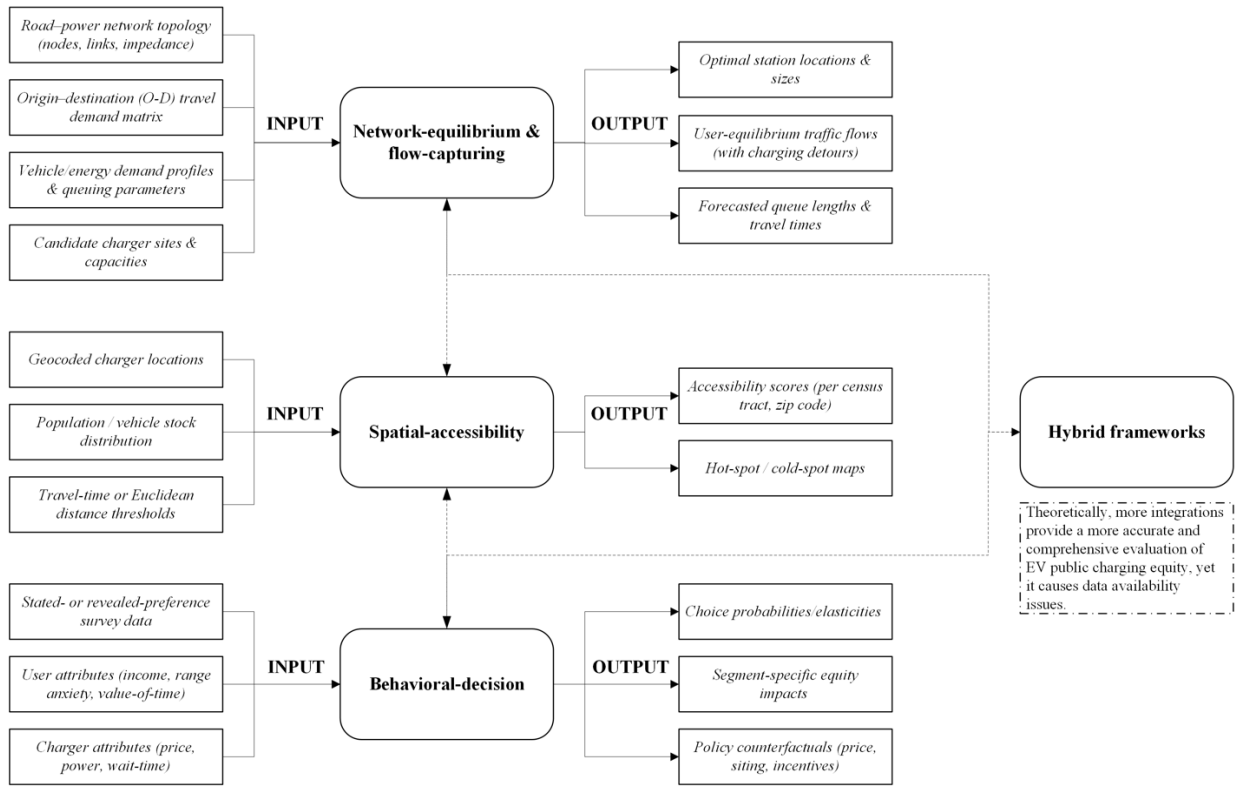

**Figure 3. Methodological Framework Synthesis**

Secondly, equity outcomes in public EV charging arise from a layered interaction of factors operating at four analytical levels and impacting four equity dimensions including SA, CB, RU, and AT. At the individual‑vehicle level, small battery capacity, high range anxiety, and utilitarian trip purposes lengthen detours and raise paid-charging frequency, worsening access and cost. Socio-economic factors such as low income, renting status, and limited digital literacy consistently increase cost burdens and erode trust, with distance effects varying by urban form. On the infrastructure level, higher-power, densely sited, and well-maintained stations reduce access and reliability penalties, although peak-price tariffs can offset these gains. Finally, system-context variables, for instance, cold climate, constrained distribution grids, and uneven policy incentives, which shape all four dimensions, especially by amplifying winter demand and voltage sag in underserved regions. Comparative evidence shows that dense urban cores enjoy short average distances but face higher tariffs and queueing, whereas rural drivers endure the opposite pattern; low-income renters bear both poor proximity and higher reliance on costly public networks, while owner-occupiers benefit from cheaper home charging despite longer travel. Cold northern cities and weak-grid regions further magnify cost and reliability gaps unless coordinated fast-charger expansion and dynamic pricing are introduced. These findings demonstrate that equitable charging provision cannot be achieved by addressing a single driver; planners could model and monitor all four dimensions and incorporate variables from each analytical level to uncover the location- and group-specific interventions that deliver the greatest equity gains.

Policy could therefore: (1) institute longitudinal, spatial-temporal monitoring of attitudes and usage; (2) apply market segmentation to remove financial and access barriers for underserved groups; and (3) tailor charger attributes, including power level, location, and pricing, to local contexts. Future work could consider refining integrated frameworks, comparing them on shared data, and fostering cross-sector collaborations to realize inclusive, equitable, and sustainable charging networks.

In conclusion, future research should focus on refining integrated analytical frameworks, conducting comparative studies across methodological approaches, and exploring advanced segmentation strategies to address diverse user needs comprehensively. Cross-sector collaborations involving policymakers, utilities, communities, and private stakeholders are crucial to promoting inclusive, equitable, and sustainable expansion of EV public charging infrastructure.